\documentclass[epj]{svjour}
\usepackage{shadow,pifont,epsfig,wrapfig,times,graphics,color}
\usepackage{latexsym}
\usepackage{graphicx, graphics, rotating, pspicture}
\usepackage{pstcol, pstricks, pst-node, pst-coil, pst-grad}

\begin{document}
\def\dsdt{$\frac{d\sigma}{dt}$}
\def\beqn{\begin{eqnarray}}
\def\eeqn{\end{eqnarray}}
\def\barr{\begin{array}}
\def\earr{\end{array}}
\def\btab{\begin{tabular}}
\def\etab{\end{tabular}}
\def\bite{\begin{itemize}}
\def\eite{\end{itemize}}
\def\bcen{\begin{center}}
\def\ecen{\end{center}}

\def\eq{\begin{equation}}
\def\ee{\end{equation}}
\def\eqa{\begin{eqnarray}}
\def\eea{\end{eqnarray}}
\def\nn{\nonumber}
\def\psmu{P^{\prime \mu}}
\def\psnu{P^{\prime \nu}}
\def\ksmu{K^{\prime \mu}}
\def\pss{P^{\prime \hspace{0.05cm}2}}
\def\psf{P^{\prime \hspace{0.05cm}4}}
\def\kdagger{K\hspace{-0.3cm}/}
\def\ndagger{N\hspace{-0.3cm}/}
\def\psdagger{p'\hspace{-0.28cm}/}
\def\epssdagger{\varepsilon'\hspace{-0.28cm}/}
\def\epsdagger{\varepsilon\hspace{-0.18cm}/}
\def\pdagger{p\hspace{-0.18cm}/}
\def\xidagger{\xi\hspace{-0.18cm}/}
\def\qsdagger{q'\hspace{-0.3cm}/}
\def\qdagger{q\hspace{-0.2cm}/}
\def\keldagger{k\hspace{-0.2cm}/}
\def\ksdagger{k'\hspace{-0.3cm}/}
\def\q2dagger{q_2\hspace{-0.35cm}/\;}
\def\qqs{q\!\cdot\!q'}
\def\lls{l\!\cdot\!l'}
\def\lp{l\!\cdot\!p}
\def\lps{l\!\cdot\!p'}
\def\lsp{l'\!\cdot\!p}
\def\lsps{l'\!\cdot\!p'}
\def\lqs{l\!\cdot\!q'}
\def\pps{p\!\cdot\!p'}
\def\psqs{p'\!\cdot\!q'}
\def\epsp{\varepsilon'\!\cdot\!p}
\def\epsps{\varepsilon'\!\cdot\!p'}
\def\epsl{\varepsilon'\!\cdot\!l}
\def\epsls{\varepsilon'\!\cdot\!l'}

\title{Electromagnetic Form Factors in the hypercentral CQM}
\author{M. De Sanctis\inst{2}  \and M.M. Giannini\inst{1} \and 
E. Santopinto \inst{1} \and A.Vassallo\inst{1} 
}                     
\institute{Universit\,a di Genova e INFN, Sezione di Genova, via Dodecaneso 33, 16142 Genova (Italy) \and INFN,Sezione di Roma1, Piazzale Aldo Moro, Roma (Italy)}
\date{Received: 05.08.2003 / Revised version: 05.08.2003}
%
\abstract{
We report on the recent results of the hypercentral Constituent
Quark Model (hCQM). The model contains a spin independent three-quark
interaction which is inspired by Lattice QCD calculations and reproduces
the average energy values of the $SU(6)$ multiplets.  The splittings are
obtained with a $SU(6)$-breaking interaction, which can include also an
isospin dependent term.
Concerning Constituent Quark models, we have shown for the first time that
the decreasing of the
ratio of the elastic form factors of the proton is due to relativistic
effects using relativistic corrections to the e.m. current and boosts.
Now the elastic nucleon form factors have been recalculated, 
using a relativistic version of the
hCQM and a relativistic quark
current showing a very detailed reproduction of all the four form factor
existing data over the complete range of $0-4$ $GeV^2$. 
Futhermore, the model has been used for predictions concerning the
electromagnetic transverse and longitudinal 
transition form factors giving a good description of the medium 
$Q^2$ behaviour. We show that the discrepancies in the
reproduction of the helicity amplitudes at low $Q^2$ are due to pion
loops. 
We have calculated the helicity amplitudes for all the 3 and 4 star 
resonances opening the
possibility of application to the evaluation of cross sections.
} 
\maketitle

\section{Introduction}
\label{intro}

In recent years much attention has been devoted to the description of the 
internal nucleon structure in terms of constituent quark degrees of
freedom. Besides the now classical Isgur-Karl model \cite{is}, the
Constituent Quark Model
has been proposed in quite different approaches: the Capstick and Isgur 
model \cite{capstick}, 
the hypercentral formulation \cite{pl} and the chiral model 
\cite{olof,ple}. In the following the hypercentral Constituent Quark Model
(hCQM), which has been used for a systematic calculation of various baryon
properties, will be briefly reviewed. The four electromagnetic elastic
 form factors of the nucleon have been recently calculated using
both a relativistic version of the hCQM and a relativistic current
and some results are presented compared with the new data. 
We have calculated in a systematic way the transverse and
longitudinal electromagnetic form factors for all the 3 and 4 star
resonances. This effort opens the
possibility to many applications for calculations of cross sections ( see
Ripani contribution for an application to the Jlab two pions
data \cite{ripani}).
Finally we will also show how it
is possible to reproduce in great detail also the behaviour at low $Q^2$ of
the helicity amplitudes for the nucleon resonances considering
chiral corrections to this model
(see L. Tiator contribution for more details on pion loop
corrections to the hCQM \cite{lothar}).
\section{The hypercentral model} 
\label{sec:1} 
\begin{figure} 
\resizebox{0.45\textwidth}{!}{%
  \includegraphics{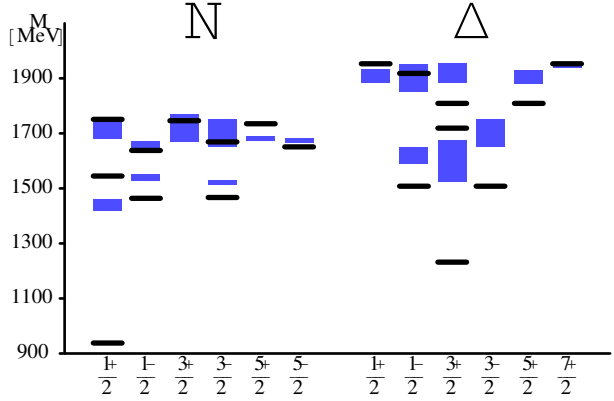}
} 
\caption{The spectrum obtained with the hypercentral model Eq.(3) and the 
parameters Eq. (4) ( full lines)), compared with the experimental data of 
PDG \cite{pdg} (grey boxes).}
\label{spettro_a_t}
\end{figure}

The experimental $4$ and $3$ star non strange resonances can be arranged
in  $SU(6)-$multiplets. This means that the quark dynamics
has a dominant $SU(6)-$ invariant part, which accounts for the average
multiplet energies. In the hCQM it is assumed to be
\cite{pl}
\begin{equation}\label{eq:pot}
V(x)= -\frac{\tau}{x}~+~\alpha x,
\end{equation}
where $x$ is the hyperradius 
\begin{equation}
x=\sqrt{{\vec{\rho}}^2+{\vec{\lambda}}^2} ~~,
\end{equation}
where $\vec{\rho}$ and $\vec{\lambda}$ are the Jacobi coordinates
describing the
internal quark motion. The dependence of the potential on the hyperangle
$\xi=arctg(\frac{{\rho}}{{\lambda}})$ has been neglected.\\
Interactions of the type linear plus Coulomb-like have been used since
long time for the meson sector, e.g. the Cornell potential. This form has been
supported by recent Lattice QCD calculations \cite{bali}.\\
In the case of
baryons a so called hypercentral approximation has been
introduced \cite{has,rich}, this approximation amounts to average 
any two-body 
potential for the three quark system over the hyperangle $\xi$ and works
quite well, specially for the lower part of the spectrum \cite{hca}. In
this respect, the hypercentral potential Eq.\ref{eq:pot} can be considered
as the hypercentral approximation of the Lattice QCD potential. On the
other hand, the hyperradius $x$ is a collective coordinate and therefore
the hypercentral potential contains also three-body effects.\\
The hypercoulomb term $1/x$ has important
features \cite{pl,sig}: it can be solved analytically and the resulting
form factors have a power-law behaviour, at variance with the widely used
harmonic oscillator; moreover, the negative parity states are exactly
degenerate with the first positive parity excitation, providing a good
starting point for the description of the spectrum.\\ 
The splittings within the multiplets are produced by a perturbative term
breaking the $SU(6)$ symmetry, 
which, as a first approximation, can be assumed to be the
standard hyperfine interaction $H_{hyp}$ \cite{is}.
The three quark hamiltonian for the hCQM is then:
\begin{equation}\label{eq:ham}
H = \frac{p_{\lambda}^2}{2m}+\frac{p_{\rho}^2}{2m}-\frac{\tau}{x}~
+~\alpha x+H_{hyp},
\end{equation}
where $m$ is the quark mass (taken equal to $1/3$ of the nucleon mass). 
The strength of the hyperfine interaction is determined in order to
reproduce the $\Delta-N$ mass difference, the remaining two free
parameters are fitted to the spectrum, reported in Fig.\ref{spettro_a_t}, 
leading to the following values:
\begin{equation}\label{eq:par}
\alpha= 1.61~fm^{-2},~~~~\tau=4.59~.
\end{equation}
Keeping these parameters fixed, the model has been applied to calculate 
various physical quantities of interest: the photocouplings \cite{aie},
the
electromagnetic transition amplitudes \cite{aie2}, the elastic nucleon
form factors \cite{mds} and the ratio between the electric and magnetic
proton form factors \cite{rap}. Some results of such parameter free
calculations are presented in the next section.

\section{The results}
\label{sec:2} 
 The electromagnetic transition amplitudes 
are defined as the matrix elements of the  
electromagnetic interaction, 
between the nucleon, $N$, and the resonance, $B$, states:
\begin{equation}\label{eq:amp}
\begin{array}{rcl}
A_{1/2}&=& \langle B, J', J'_{z}=\frac{1}{2}\ | H_{em}^t| N, J~=~
\frac{1}{2}, J_{z}= -\frac{1}{2}\
\rangle\zeta\\
& & \\
A_{3/2}&=& \langle B, J', J'_{z}=\frac{3}{2}\ | H_{em}^t| N, J~=~
\frac{1}{2}, J_{z}= \frac{1}{2}~\
\rangle~\zeta~\\
& & \\
S_{1/2}&=& \langle B, J', J'_{z}=\frac{1}{2}\ | H_{em}^l| N, J~=~
\frac{1}{2}, J_{z}= \frac{1}{2}~\
\rangle~\zeta~\\ 
\end{array}
\end{equation}
where $\zeta$ is the sign of the N$\pi$ amplitude.

\begin{figure}[h] 
\resizebox{0.50\textwidth}{!}{%
\includegraphics[angle=90]{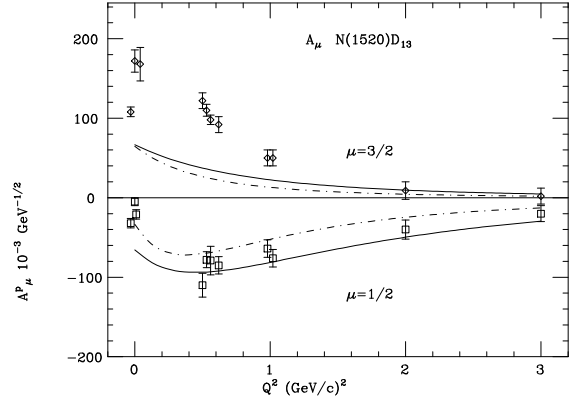} 
} 
\caption{
The helicity amplitudes for the $D_{13}(1520)$ resonance,
calculated with the hCQM of Eqs. (3) and (4) (full curve, \cite{aie2}).
The dashed curve is obtained with the analytical version of the hCQM
(\cite{sig}), where the behaviour of the quark wave function is determined
mainly by the hypercoulomb potential. The data are from the compilation of
ref. \cite{burk}
}
\label{d13}       
\end{figure} 

The proton photocouplings of the hCQM \cite{aie} (Eq.~(\ref{eq:amp})
calculated  at the photon point), in comparison with other
calculations \cite{ki,cap}, have the same overall behaviour,
having the same SU(6)
structure in common, but in many cases they all show a lack of strength.\\
\begin{figure}[hb] 
\resizebox{0.45\textwidth}{!}{%
\includegraphics{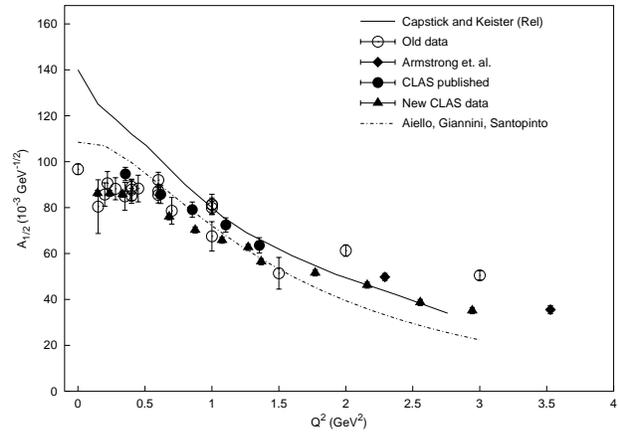} 
} 
\caption{
The helicity amplitudes for the $S_{11}(1535)$ resonance,
calculated with the hCQM of Eqs. (3) and (4) (dotted curve, \cite{aie2}) 
and the model of ref.
\cite{cap} (full curve). 
The data are taken from the compilation of ref. \cite{burk2}
}
\label{s11}       
\end{figure} 
Taking into account the $Q^2$ behaviour of the transition matrix elements 
of Eq.~(\ref{eq:amp}), one can calculate the hCQM helicity amplitudes in
the Breit frame \cite{aie2}. The hCQM results for the $D_{13}(1520)$ and
the $S_{11}(1535)$ resonances \cite{aie2} are given in Fig.\ref{d13} and 
\ref{s11},
respectively. The agreement in the case of the $S_{11}$ is remarkable, the
more so since the hCQM curve has been published three years in advance
with respect to the recent TJNAF data \cite{dytman}.
We have completed our program in order to calculate
in a systematic way the helicity amplitudes,  transverse and longitudinal 
ones, for all the 3 and 4 star resonances 
( the results are at disposal under request) \cite{long}. 
In general the $Q^2$ behaviour is reproduced, except for
discrepancies at small $Q^2$, especially in the
$A^{p}_{3/2}$ amplitude of the transition to the $D_{13}(1520)$ state. 
These discrepancies, as the ones observed in the photocouplings, can be
ascribed either to the non-relativistic character of
the model or to the lack of explicit quark-antiquark configurations, 
which may be important at low $Q^{2}$ .  The kinematical relativistic
corrections at the level of boosting the nucleon  and the resonance
states to a common frame are not responsible for these discrepancies,  as
we have demonstrated in Ref.~\cite{mds2}. Similar results are obtained for 
the other negative parity resonances \cite{aie2}. 

It should be mentioned that the r.m.s. radius of the proton corresponding 
to the parameters of Eq.~(\ref{eq:par}) is $0.48$ fm, which is the same 
value obtained in \cite{cko} in order to reproduce the $D_{13}$
photocoupling. Therefore the missing strength at low $Q^2$ can be ascribed
to the lack of quark-antiquark effects, probably more important in the
outer region of the nucleon.

\begin{figure}[hb] 
\resizebox{0.45\textwidth}{!}{%
\includegraphics {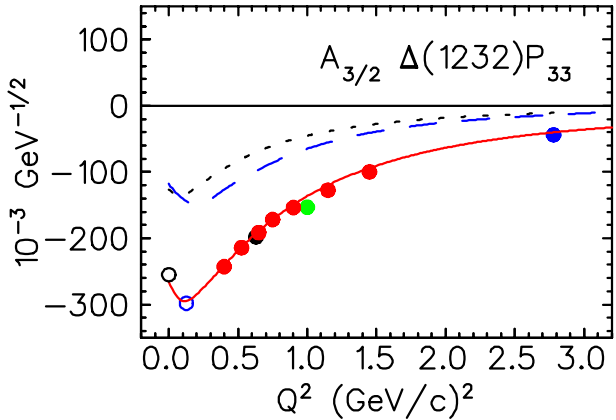}
}
\caption{
The transverse $A_{3/2}$ 
helicity amplitude for the $\Delta(1232)$ resonance. 
The dotted line corresponds to the hCQM results, the dashed line to the pion 
loop contributions and the full line to a fit of the existing data  
(see L.Tiator contribution for a complete explanation).
}
\label{helamp}       
\end{figure}

For example,
for the Delta resonance the contribution of the pion cloud is 
very important  \cite{lothar}. 
For the transverse amplitudes $A_{1/2}$ and $A_{3/2}$ it is 
about $50\%$ at low $Q^2$ and for the longitudinal amplitude as well as for 
the electric amplitude the pion cloud is absolutely dominant.\\
In Fig. \ref{helamp} we show for the $A_{3/2}$ amplitude that only the sum 
of both contribution will get close to the empirical results.
%
\section{The isospin dependence}
In the chiral Constituent Quark Model \cite{olof,ple}, the non
confining part of the   
potential is provided by the interaction with the Goldstone bosons,
giving rise to a spin- and flavour-dependent part, which is crucial in
this approach for the description of the lower part of the spectrum.
More generally, one can expect that the quark-antiquark pair production 
can lead to an effective residual quark interaction containing an isospin
(flavour) dependent term.
\begin{figure} 
\resizebox{0.45\textwidth}{!}{%
\includegraphics{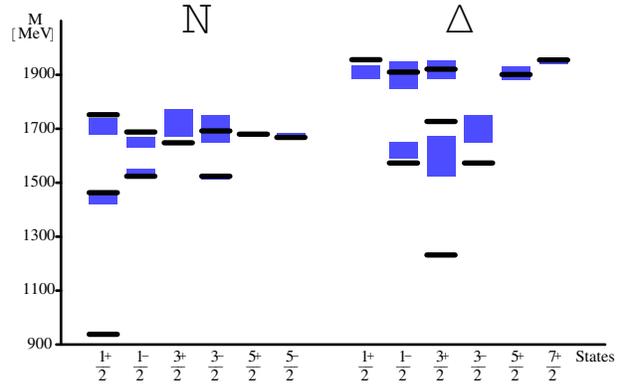} 
} 
\caption{
The spectrum obtained with the hypercentral model
containing isospin dependent terms Eq. (7) \cite{iso} (full lines)),
compared with the
experimental data of PDG \cite{pdg} (grey boxes)
}
\label{iso}       
\end{figure} 
Therefore, we have introduced isospin dependent terms in the hCQM 
hamiltonian. The complete interaction used is given by \cite{iso}
\begin{equation}\label{tot}
H_{int}~=~V(x) +H_{\mathrm{S}} +H_{\mathrm{I}} +H_{\mathrm{SI}}~,
\end{equation}
where $V(x)$ is the linear plus hypercoulomb SU(6)-invariant potential of Eq.
\ref{eq:pot}, while the remaining terms are the residual SU(6)-breaking
interaction, responsible for the splittings within the multiplets. 
${H}_{\mathrm{S}}$ is a smeared standard hyperfine term,  
${H}_{\mathrm{I}}$ is isospin dependent and  ${H}_{\mathrm{SI}}$ 
spin-isospin dependent.
The resulting spectrum for the 3 and 4 star resonances is shown in Fig. 
\ref{iso} \cite{iso}. The contribution of the
hyperfine interaction to the $N-\Delta$ mass difference is in this case 
only about $35\%$, while the remaining splitting comes from the
spin-isospin term, $(50\%)$, and from the isospin one, $(15\%)$.
It should be noted that the position of the Roper and the negative 
parity states is well reproduced.

\section{Relativity}

The relativistic effects that one can introduce starting from a non
relativistic quark model are:
a) the relativistic kinetic energy;
b) the boosts from the rest frames of the initial and final baryon to a 
common (say the Breit) frame;
c) a relativistic quark current. All these features are not equivalent to 
a fully relativistic dynamics, which is still beyond the present 
capabilities of the various models.

The potential of Eq.\ref{eq:pot} has been refitted using the correct 
relativistic kinetic energy
\begin{equation}\label{eq:hrel}
H_{rel} = \sum_{i=1}^3 \sqrt{p_{i}^2+m^2}-\frac{\tau}{x}~
+~\alpha x+H_{hyp}.
\end{equation}
 The resulting spectrum is not much different from the non relativistic
one and the parameters of the potential are only slightly modified.

The boosts and a relativistic quark current expanded up to lowest order 
in the quark momenta has been used both for the elastic form factors of
the nucleon \cite{mds} and the helicity amplitudes \cite{mds2}. In the
latter case, as already mentioned, the relativistic effects are quite
small and do not alter the agreement with data discussed previously. For
the elastic form factors, the relativistic effects are quite strong and
bring the theoretical curves much closer to the data; in any case they are
responsible for the decrease of the ratio between the electric and magnetic
proton form factors, as it has been shown for the first time in Ref.~ 
\cite{rap}, in qualitative agreement with the recent 
Jlab data \cite{ped}.

A relativistic quark current, with no expansion in the quark momenta, and 
the boosts to the Breit frame have been applied to the calculation of the
elastic form factors in the relativistic version of the hCQM
Eq. (\ref{eq:hrel}) \cite{mds3}.
The resulting theoretical form factors of the proton, calculated, it
should be stressed, without free parameters and assuming pointlike quarks,
are good 
with some discrepancies at low $Q^2$, which, as discussed
previously, can be attributed to the lacking of the quark-antiquark pair 
effects. Concerning the ratio between the electric and magnetic proton
form factors 
the deviation from unity reaches almost the $50\%$ level, not far
from the new TJNAF data \cite{gay}. 
\begin{figure}[hb]
\resizebox{0.35\textwidth}{!}{%
\begin{pspicture}(0,0)(10,9) 
\put(0.5,0.5){\includegraphics{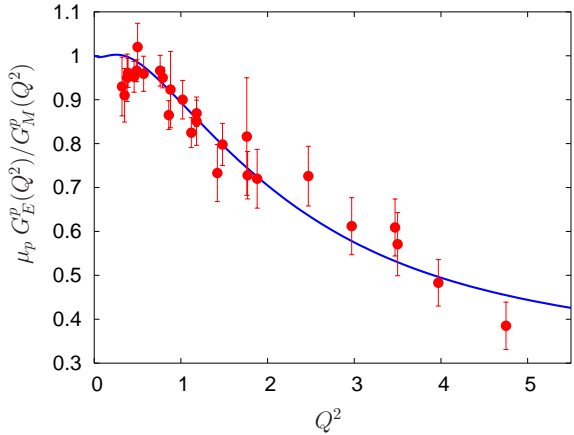}}
\put(0.5,3.5){\Large \rotatebox{90}{$\mu_p~G_E^p(Q^2)/G_M^p(Q^2)$}}
\put(7.0,0.0){\Large $Q^2$}
\end{pspicture} 
} 
\caption{The ratio between the electric and magnetic proton form
factors, calculated with the relativistic hCQM of eq. (8),
a relativistic current and a small constituent quark form factor \cite{mds3},
 compared with the TJNAF data 
\cite{ped,gay}}
\label{rap}       
\end{figure} 
Nevertheless to obtain a full 
description of the existing data on the elastic form factors one still has to 
take into account the $q\bar{q}$ degrees of freedom. A way to 
describe effectively the effect of these extra degrees of freedom is to 
introduce constituent quark form factors. 
Adding the effect of small contituent quark form factor to the hCQM results, 
the curve shown in Fig. \ref{rap} is obtained.  

\section{Conclusions}

\begin{figure} 
\resizebox{0.45\textwidth}{!}{%
\includegraphics{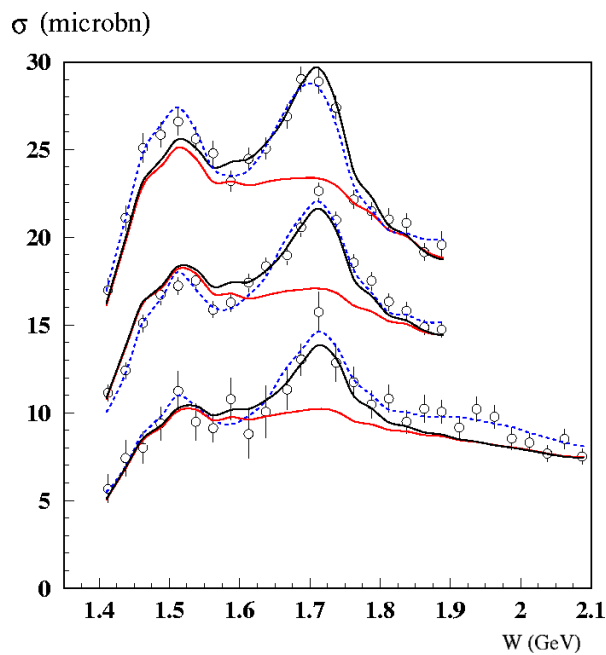}}  
\caption{Comparison of the new CLAS data for the two pion total cross section 
with the calculation based on the electromagnetic helicity amplitudes 
from the hCQM, 
transverse and longitudinal ones, the solid line is the result 
obtained including 
also the new state found by their previous analysis. The  explanation of 
this picture and of the work can be found on Ripani's contribution.  } 
\label{ripani}       
\end{figure}

The hCQM is a generalization to the baryon sector of the widely used
quark-antiquark
potential containing a coulomb plus a linear confining term. The three free
parameters have been adjusted to fit the spectrum \cite{pl} and then the 
model has been used for a systematic calculation of various physical
quantities: the photocouplings \cite{aie}, the helicity amplitudes for the
electromagnetic excitation of negative parity baryon resonances
\cite{aie2,mds2,long}, the elastic form factors of the nucleon
\cite{mds,mds3}
and the ratio between the electric and magnetic proton form
factors \cite{rap,mds3}. The agreement with data is quite good, specially
for the helicity amplitudes, which are reproduced in the medium-high $Q^2$
behaviour, leaving some discrepancies at low (or zero) $Q^2$, where the
lacking quark-antiquark contributions are expected to be effective. It
should be noted that the hypercoulomb term in the potential is the main
responsible of such an agreement \cite{sig}, while for the
spectrum a further fundamental aspect is provided by the isospin dependent
interactions \cite{iso}. We have completed our program calculating the 
transverse and longitudinal helicity amplitudes for all the resonances 
\cite{long}, opening in this way the possibility of
applications to the calculation of cross sections, as for example in the
two pion case \cite{ripani} (see Fig. \ref{ripani}).
Finally, we have calculated the chiral corrections to our helicity
amplitudes showing an impressive reproduction of all the existing data
\cite{lothar}.

%

%
 
\end{document}